\documentclass[aps,amssymb,pra,twocolumn]{revtex4} \fussy

\usepackage{ifpdf} 
\ifpdf

\usepackage[pdftex]{graphicx,color} 
\usepackage{epstopdf}  
\else

\fi

\usepackage{epsfig,graphicx}
\usepackage{float}

\newcommand{\figwidth}{0.45\textwidth}

\begin{document}

\title{Orientation-dependent spontaneous emission rates of a two-level quantum emitter in any nanophotonic environment}

\author{Willem L. Vos$^{1,2}$}
\altaffiliation{Email: w.l.vos@amolf.nl, web: www.photonicbandgaps.com}

\author{A. Femius Koenderink$^{1}$}
\altaffiliation{Email: f.koenderink@amolf.nl, web: www.koenderink.info}

\author{Ivan S. Nikolaev$^{1}$}
\altaffiliation{Now at ASML, Veldhoven, the Netherlands.}

\affiliation{$^1$Center for Nanophotonics, FOM Institute for Atomic and Molecular Physics AMOLF, Amsterdam, The Netherlands}

\affiliation{$^2$Complex Photonic Systems (COPS), MESA$^+$ Institute for Nanotechnology, University of Twente, The Netherlands}

\date{Published as Phys. Rev. A 80, 053802, 2009.}

\begin{abstract}
We study theoretically the spontaneous emission rate of a two-level quantum emitter in any nanophotonic system.
We derive a general representation of the rate on the orientation of the transition dipole by only invoking symmetry of the Green function.
The rate depends quadratically on orientation and is determined by rates along three principal axes, which greatly simplifies visualization:
Emission-rate surfaces provide insight on how preferred orientations for enhancement (or inhibition) depend on emission frequency and location, as shown for a mirror, a plasmonic sphere, or a photonic bandgap crystal.
Moreover, insight is provided on novel means to "switch" the emission rates by actively controlling the orientation of the emitters' transition dipole.

\end{abstract}

\maketitle

\section{Introduction}
It is well-known that the characteristics of spontaneously emitted light depend strongly on the environment of the light source
\cite{Drexhage70,Kleppner81,Haroche92,Novotny06}.
According to quantum electrodynamics, the emission rate of a two-level quantum emitter, described by Fermi's golden rule, is generally factorized into a part describing the sources' intrinsic quantum properties and another part describing the influence of the environment on the light field.
Currently, there are many efforts to control the emission rate of quantum emitters by optimizing the nanoscale environment by, \emph{e.g.}, reflecting interfaces \cite{Drexhage70,Snoeks95}, microcavities \cite{Gerard98,Englund05}, photonic crystals \cite{Yab87,Sprik96,Lodahl04,Busch&John98}, or plasmonic nanoantennae \cite{Farahani05,Muskens07,Tam08}.
Control of spontaneous emission is notably relevant to applications, including single-photon sources for quantum information, miniature lasers and light-emitting diodes, and solar energy harvesting~\cite{Graetzel01,San02,Park04}.

\begin{figure}
\includegraphics[width=\columnwidth]{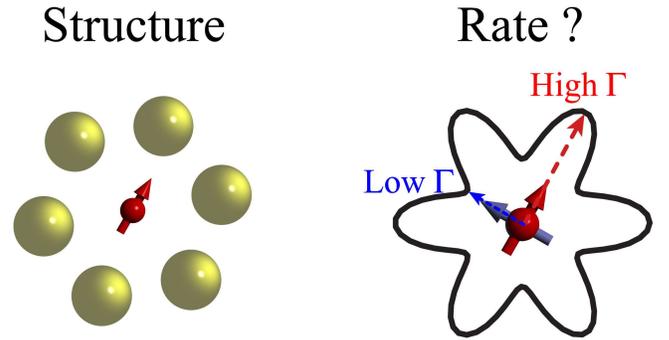}
\caption{\label{5cartoon} (color online) Drawing of a two-level
quantum emitter embedded  in an arbitrary nanophotonic system, here
depicted as a cluster of 6 scatterers. If the emission rate were to
mimic the symmetry of the system, one would here expect an
emission-rate surface with a 6-fold symmetry. Our analysis reveals,
however, that these surfaces take on only specific shapes determined
by the symmetry of the Green dyadic. The symmetry analysis allows
one to conclude without any calculation that the rate is identical
for all dipole orientations in the plane of the 6 scatterers. }
\end{figure}

The effect of the environment of a source on its emission rate is described by the local density of optical states (LDOS)
\cite{Sprik96,Busch&John98,Novotny06}. The LDOS counts the number of photon modes available for emission, and it is interpreted as the density of vacuum fluctuations. In many experimentally relevant cases, it is theoretically known that emission rates strongly differ for various orientations of the transition dipole moment see, \emph{e.g.},~\cite{Novotny06,Kuhn2008}.
Thus, the widely pursued control of position and frequency leaves a large uncertainty in the emission rate~\cite{Sprik96}.
To date, no clear picture has emerged of the general characteristics of the orientation dependence.
It is an open question whether the behavior mimics the local symmetry around the emitter, see Fig.~\ref{5cartoon}, or whether any generic dependence exists at all.

%
Therefore, we present fundamental insights in the complex dependence of the
emission rates of a quantum emitter on the orientation of its dipole moment. Our general, yet simple theoretical
analysis only invokes the symmetry of the Green function and provides a complete classification of the orientation-dependences
that the emission rate can assume in any nanophotonic system.
This classification leads to an intuitive visualization that is based on only a few clearly defined physical parameters, as shown by examples of an emitter near a mirror, a plasmonic sphere, or in a 3D photonic bandgap crystal.
From our analysis, we conclude that control over the \textit{orientation} of the transition
dipole moment opens novel applications: If one can tune the orientation of an emitter, one can "switch" emission from inhibited to enhanced and \textit{vice versa}. In the field of quantum information \cite{Kimble08}, atomic qubits that fly by
nanophotonic systems could acquire controllable phase shifts by tuning their orientation relative to the principal axes.

\section{Theory}
\subsection{Derivation of emission-rate surface}
The rate of spontaneous emission $\Gamma$ of a two-level dipolar quantum emitter in the weak-coupling approximation is equal to
\cite{Sprik96,Busch&John98,Novotny06}:
\begin{equation}\label{rate_LDOS}
\Gamma(\omega,\mathbf{r},\mathbf{e}_d) = \frac{\pi d^2
\omega}{\hbar \epsilon_0} N(\mathbf{r},\omega,\mathbf{e}_d),
\end{equation}
with $\omega$ the emission frequency, $\mathbf{r}$ the source's
position, $\mathbf{e}_d$ the dipole orientation, $d$ the modulus
of the matrix element of the transition dipole moment.
$N(\mathbf{r},\omega,\mathbf{e}_d)$ is the local density of
optical states (LDOS) that equals:
\begin{equation}\label{LDOS2Green}
N(\omega,\mathbf{r},\mathbf{e}_d)= \frac{6 \omega}{\pi c^2}
(\mathbf{e}_d^T \cdot
\mathrm{Im}(\mathbb{G}(\mathbf{r},\mathbf{r},\omega)) \cdot
\mathbf{e}_d),
\end{equation}
\noindent with $\mathbb{G}(\mathbf{r},\mathbf{r},\omega)$ the Green
dyadic \cite{Novotny06}. Eq.~(\ref{rate_LDOS}) reveals the well-known fact that the emission rate depends on the frequency and
the position of the emitter. As is well known, Eq.~(\ref{LDOS2Green}) is also applicable to emission dynamics
inside dissipative optical media. In such media, the imaginary part of the Green dyadic describes the \emph{total} decay rate, \emph{i.e.}, the sum of the radiative decay rate and the rate of quenching induced by the environment. Hence, the results in this paper carry over straightaway to the decay dynamics of dipoles emitters in dissipative nanophotonic environments.

\begin{figure*}
\includegraphics[width=\textwidth]{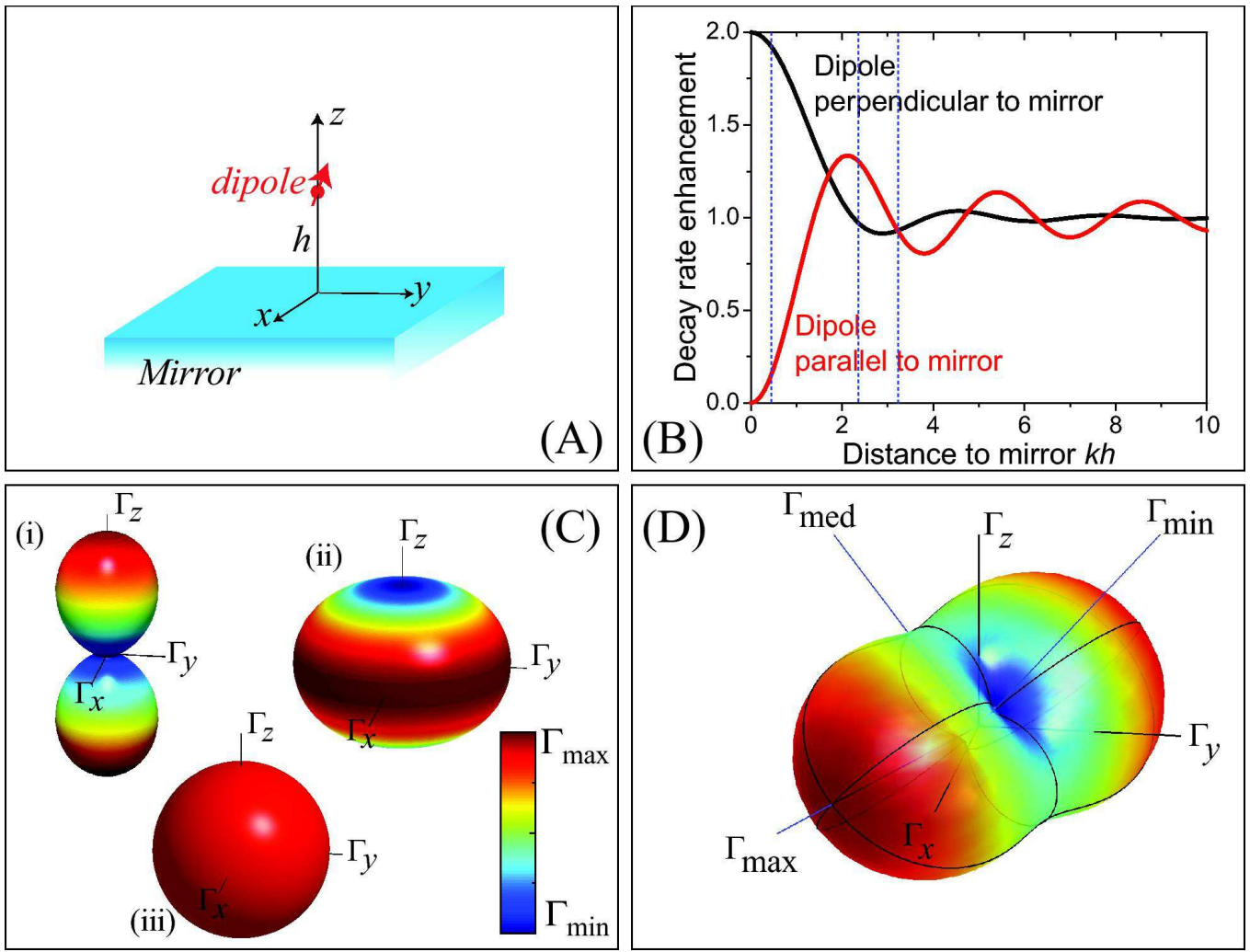}
\caption{\label{peanut1} (color) \textbf{(A)} Drawing of a two-level
quantum emitter at distance $h$ above a mirror. \textbf{(B)}
Emission rate versus scaled distance (wave vector times distance
$kh$) for a dipole perpendicular and parallel to a perfect
mirror~\cite{Haroche92}. \textbf{(C)} Three-dimensional surfaces
representing the orientation dependent spontaneous emission rate in
real space. (i) One maximal emission and two equal minimal rates
give a peanut-shape (at $kh = 0.4$ in (B)). (ii) Two equal maximal
rates and one minimal rate give an oblate spheroid ($kh = 2.3$ in
(B)). (iii) Three equal maximal rates give a sphere ($kh = 3.2$ in
(B)). \textbf{(D)} Most general shape when all principal rates are
different
($\Gamma_{\mathrm{max}}>\Gamma_{\mathrm{med}}>\Gamma_{\mathrm{min}}$)
and the principal axes are rotated from the $(x,y,z)$-axes. Color
scales are linear from $\Gamma_{\mathrm{min}}$ to
$\Gamma_{\mathrm{max}}$ (colorbar in (C)).}
\end{figure*}

A didactic example to illustrate the dependence of emission rates on
frequency, position and dipole orientation is that of a source near
a perfect mirror, see Fig.~\ref{peanut1}(A), which can be understood
from image dipole analysis \cite{Haroche92,Drexhage70}. The rate
depends strongly on the dipole orientation $\mathbf{e}_d$: at small
distances a dipole parallel to the mirror has a vanishing emission
rate, which can be interpreted as due to destructive interference of
the dipole with its oppositely oriented image. In contrast, a dipole
perpendicular to the mirror has twice the unperturbed rate owing to
constructive interference, as shown in Fig.~\ref{peanut1}(B).
Clearly, the symmetry of this particular geometry implies that the
parallel and perpendicular dipole orientations are `principal'
orientations along which the maximum and minimum rates are attained.
At intermediate orientations the rate is a weighted average of the
two rates.

The main result of our paper is that the rate \emph{always} depends on orientation via a quadratic form with three perpendicular principal axes, as will now be proven:
On account of reciprocity, the Green dyadic is equal to its transpose upon exchanging the coordinates. Hence
\begin{equation}\label{GreenSym}
\mathrm{Im}(\mathbb{G}(\mathbf{r},\mathbf{r'},\omega))^{T}=\mathrm{Im}(\mathbb{G}(\mathbf{r'},\mathbf{r},\omega)).
\end{equation}
Furthermore, the imaginary part of the Green dyadic is real.
Therefore, the imaginary part of the Green dyadic in
Eq.~(\ref{LDOS2Green}) is a real and symmetric $3\times 3$ matrix.
Consequently, at each frequency $\omega$ and spatial position
$\mathbf{r}$, the imaginary part of the Green dyadic can always be
diagonalized, and has 3 eigenvalues ($g_{1}$, $g_{2}$, $g_{3}$) that
correspond to three orthogonal eigenvectors. Since the eigenvalues
can be ordered by magnitude, we relabel the eigenvalues and the
concomitant main axes as
$\{\mathbf{v}_\mathrm{min},\mathbf{v}_\mathrm{med},\mathbf{v}_\mathrm{max}\}$.
This basis corresponds to three perpendicular principal dipole
orientations that vary with dipole location $\mathbf{r}$ and
frequency $\omega$. In this orthonormal basis we express the dipole
orientation unit vector $\mathbf{e}_d$ as:
\begin{equation}\label{unitdipolebasis}
\mathbf{e}_d = \beta_{1} \mathbf{v}_\mathrm{min} + \beta_{2} \mathbf{v}_\mathrm{med} + \beta_{3} \mathbf{v}_\mathrm{max},
\end{equation}
\noindent where $\beta_{i}$ are coefficients that are  constrained
through $\beta_{1}^{2}+\beta_{2}^{2}+\beta_{3}^{2}=1$ to lie on a
unit sphere, since $\|\mathbf{e}_d\|=1$. Clearly, the coefficients
$\beta_{i}$ are functions of the dipole orientation:
$\beta_{i}=\beta_{i} (\mathbf{e}_d)=
\mathbf{e}_d^T\cdot\mathbf{v}_i$.

Using Eqs.~(\ref{rate_LDOS}, \ref{LDOS2Green}),  the emission rate
$\Gamma$ can be expressed in emission rate coefficients
$\Gamma_{i}$, which are the rates for dipole orientations parallel
to the principal axes $\mathbf{v}_\mathrm{i}$, leading to:
\begin{equation}\label{decayratebasis}
\Gamma(\mathbf{e}_d) = \beta_{1}^2(\mathbf{e}_d) \Gamma_{\mathrm{min}} + \beta_{2}^2(\mathbf{e}_d) \Gamma_{\mathrm{med}} + \beta_{3}^2(\mathbf{e}_d) \Gamma_{\mathrm{max}}.
\end{equation}
\noindent Equation~(\ref{decayratebasis})  describes the
emission-rate surface as a function of dipole orientation
$\Gamma(\mathbf{e}_d)$, which is a central result of our work. The
emission rate coefficients $\Gamma_{i}$ are equal to:
\begin{equation}\label{ratecomponent}
\Gamma_{i} = \frac{\pi d^2 \omega}{\hbar \epsilon_0} \frac{6
\omega}{\pi c^2} (\mathbf{v}_\mathrm{i}^T \cdot
\mathrm{Im}(\mathbb{G}) \cdot \mathbf{v}_\mathrm{i}) = \frac{\pi
d^2 \omega}{\hbar \epsilon_0} \frac{6 \omega}{\pi c^2}g_{i},
\end{equation}
\noindent and are via $\mathbb{G}$ functions of the frequency and
the dipoles' position: $\Gamma_{i}=\Gamma_{i}(\omega,\mathbf{r})$.
Assuming known principal rates $\Gamma_{i}$, the emission-rate surface $\Gamma(\mathbf{e}_d)$ is always a quadratic form on the unit sphere. Moreover only quadratic forms of signature $s=\sum(\mbox{sign}(\Gamma_i))=3$ can occur~\cite{mathencyc}, since emission rates are physically constrained to be positive for all orientations. Therefore, polar plots of the rate versus dipole orientation - henceforth called \emph{emission-rate surface} - take on only specific shapes classified by the ratios of $\Gamma_{\mathrm{min}},\Gamma_{\mathrm{med}}$, $\Gamma_{\mathrm{max}}$, with three perpendicular symmetry axes,
regardless of the nanophotonic system. We remark that while
Eq.~(\ref{decayratebasis}) may appear as the defining equation  of
an ellipsoid, the emission-rate surface is \emph{not} an ellipsoid
since the problem is not about calculating a level surface  of Eq.
(\ref{decayratebasis}), which would be equivalent to constraining
$\beta_{i}$ to yield a fixed $\Gamma$ in Eq. (\ref{decayratebasis}),
rather than constraining $\beta_{i}$ to the unit sphere. Our result
that emission surfaces are always necessarily quadratic forms defies
the intuition (as sketched in Fig.~\ref{5cartoon}) that emission
rates inherit the symmetry of the nanophotonic system.

Regarding the assumptions we require to arrive at the quadratic form
for the emission rate surfaces, we note that we have assumed real
dipole moment in Eq.~(\ref{LDOS2Green}) (following
Ref.~\cite{Novotny06}) and that we used reciprocity to ensure real
and symmetric $\mbox{Im}(\mathbb{G}(\mathbf{r},\mathbf{r}))$. In
case of reciprocal media it is easy to show that our results are
also valid for complex transition dipole moments, and not just for
real dipole moments. Furthermore, if we assume a  real dipole
moment, it appears that our results are also valid for metamaterials
that violate reciprocity, i.e., in case
$\mbox{Im}(\mathbb{G}(\mathbf{r},\mathbf{r}))$ is not symmetric or
even not diagonalizable. Since
$\mbox{Im}(\mathbb{G}(\mathbf{r},\mathbf{r}))$ is still real it will
nonetheless give rise to a quadratic form that can be transformed to
a principal axis system~\cite{mathencyc}. The physical requirement
that rates are positive for all dipole orientations furthermore
ensures that the signature of the quadratic form remains $3$ even in
the nonreciprocal case.

\subsection{Generic shapes of the emission-rate surface}
Figure~\ref{peanut1}(C,D) categorizes all possible shapes of the emission rate polar plot. Fig.~\ref{peanut1}(C) is relevant for the mirror, with principal axes parallel ($x,y$, degenerate) and perpendicular ($z$)
to the interface. Fig.~\ref{peanut1}(C(i)) shows the emission-rate
surface for the case where emission is enhanced along a single
dipole orientation $\Gamma_{\mathrm{max}} \gg
\Gamma_{\mathrm{min}}=\Gamma_{\mathrm{med}}$. This situation appears
at a reduced distance $kh = 0.4$ close to the mirror. Here, the
emission-rate surface looks like a highly anisotropic peanut,
constricted to a radius $\Gamma_{\mathrm{min}}$ in the $x,y$-plane,
and extending to $\Gamma_{\mathrm{max}}$ along the $z$-axis.
Fig.~\ref{peanut1}(C(ii)) shows the orientation dependent emission
rate for a single inhibited axis with $\Gamma_{\mathrm{min}} \leq
\Gamma_{\mathrm{med}} = \Gamma_{\mathrm{max}}$, at $kh=2.3$ near a
mirror. Qualitatively, the emission-rate surface resembles an oblate
spheroid; when the minimum rate is much less than the other two
rates (see Fig.~\ref{GammaPlasmonSphere} below), the surface
develops a concave indentation with a donut-like shape.
Fig.~\ref{peanut1}(C(iii)) shows the emission-rate surface when the
rate is equal along all three main axes ($kh = 3.2$). The
emission-rate surface is simply a sphere, as it is in any isotropic
homogenous medium.

Fig.~\ref{peanut1}(D) shows the emission-rate surface for the \textit{most general} case when i) the rates along the main axes are
all different ($\Gamma_{\mathrm{min}} < \Gamma_{\mathrm{med}} < \Gamma_{\mathrm{max}}$), and ii) the principal axes
$\mathbf{v}_{\mathrm{min,med,max}}$ have an arbitrary orientation with respect to the laboratory frame. Clearly, the emission rate is not extremal for a dipole parallel to any of the $(x,y,z)$-axes.
An important feature of the emission-rate surfaces is that they allow for an easy inspection of both the anisotropy of the emission rates, and of the favorable dipole orientations compared to the usual $(x,y,z)$-axes in real space.

\section{Efficient method to calculate emission-rate surfaces}
In many cases of practical interest, neither the Green's function
$\mathbb{G}$ nor the principal axes $\{\mathbf{v}_\mathrm{i}\}$ are
\textit{a-priori} known. Often algorithms based on a summation over
all photon modes are used that only yield the rate $\Gamma$ for
target orientations $\mathbf{e}_d$ chosen as \emph{a priori} input.
Reconstructing emission-rate surfaces as in Fig.~\ref{peanut1} by a
dense sampling of orientations is not viable with such algorithms,
due to prohibitive computation times. A poignant example is the
calculation of emission rates in photonic crystals that requires a
summation over up to $10^6$ Bloch modes, the calculation of each of
which requires diagonalization of a $10^3\times 10^3$ matrix, even
for a single dipole orientation~\cite{Busch&John98,Nikolaev09}. A
popular alternative method that can conveniently yield the emission
rate for a single orientation is the finite difference time domain
(FDTD) simulation method~\cite{FDTD}. However, it appears difficult
to calculate off-diagonal elements of the Green tensor. Since the
various field components are not calculated on identical grid
points, FDTD does not truly yield a Green dyadic on a well-defined
position $\mathbf{r}$. Hence, even if an algorithm is known to
calculate rates at fixed orientations, it is unclear how to find the
principal axes and rates, since $\mbox{Im}(\mathbb{G})$ is simply
not available for diagonalization. In view of the computational cost
of evaluating the radiative rate at a single dipole orientation, the
main problem is to find out for how many and for which orientations
the emission rate must be calculated to completely and exactly
characterize the emission-rate surfaces. Here we describe an
efficient method to find principal emission rates and orientations
by evaluating the LDOS at the \emph{least possible number\/} of
input orientations.

We use the well-known fact that any function on the unit sphere  is
conveniently expanded in spherical harmonics $Y_{lm}(\theta,\phi) =
P_{lm}(\cos(\theta))e^{im\phi}$. Since the emission surface is a
quadratic form, we can apply the well-known fact that all quadratic
forms on the unit sphere can be represented exactly by an expansion
containing only terms up to $l = 2$, so that
\begin{equation}\label{rateLegendre}
\Gamma(\mathbf{e}_d) = \sum_{l=0}^{2} \sum_{m=-l}^{l} a_{lm} P_{lm}
(\cos(\theta))e^{im\phi}
\end{equation}
An easy proof that no terms beyond $l=2$ are needed is obtained by
expressing the spherical harmonics in terms of cartesian
coordinates, rather than polar coordinates on the unit
sphere~\cite{mathencyc}, or conversely by expressing the
coefficients $\beta_{i}$ in terms of polar coordinates relative to
the $\{\mathbf{v}_\mathrm{i}\}$ axis system. This substition leads
to a trigonometric expansion for $\Gamma(\mathbf{e}_d)$ with terms
that are  quadratic in cosines and sines of $\theta$ and of $\phi$,
see Appendix~\ref{Averages}, Eq.~(\ref{GammaPrincip}).

\begin{figure}
\includegraphics[width=\figwidth]{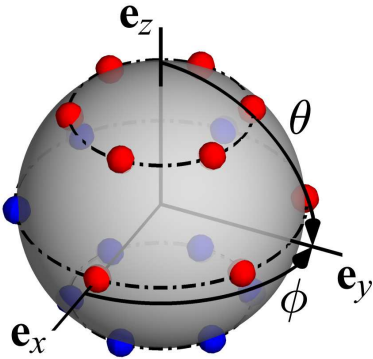}
\caption{\label{specialpoints} (color online) Special orientations, \emph{i.e.},
points on the unit sphere, for which the decay rate needs to be
calculated in order to fully reconstruct emission-rate surfaces. The
blue and red points together are the roots of $l=3$ spherical
harmonics, corresponding to $\phi=m\pi/3$ ($m=0,1,2,\ldots 5$) and
$\theta=\arccos(\sqrt{3/5}),\pi/2,-\arccos(\sqrt{3/5})$. Due to
inversion symmetry, rates are equal for antipodal orientations. This
makes calculations for half the points (e.g., the blue ones)
superfluous, leaving 9 distinct orientations (red) for which rates
must be calculated in order to find principal rates and orientations. Note that these points do not have equal weights in
Eq.~(\ref{acoefficients2}) (weight is $8/9$ for points on the
equator and $5/9$ for other points).}
\end{figure}

The expansion coefficients for the spherical harmonic expansion are given by inner products
\begin{equation}\label{acoefficients}
a_{lm} = \langle \Gamma(\mathbf{e}_d), Y_{lm} \rangle =
\int_0^{2\pi} d\phi \int_{0}^\pi d\theta~\Gamma(\mathbf{e}_d) Y_{lm}
(\theta,\phi)\sin(\theta),
\end{equation}
similar to the coefficients appearing in discrete Fourier
transformations,  but now for transformation on the unit sphere.
Mohlenkamp has developed a fast Fourier transform method to
calculate the coefficients numerically~\cite{Mohlenkamp99}, which
requires a sampling of rates $\Gamma$ at a discrete set of
orientations, similar to the numerical evaluation of discrete
Fourier coefficients by the sampling of a periodic function on a
discrete set of points. In this approach, the integral
expression~(\ref{acoefficients}) for the expansion coefficients for
expanding a function $f$ is replaced by a discrete weighted sum:
\begin{equation}\label{acoefficients2}
\hat{a}_{lm} = \sum_{k} w_{k} f(\theta_{k},\phi_{k})
Y_{lm}(\theta_{k},\phi_{k}) \sin(\theta_{k}),
\end{equation}
\noindent where $k$ runs over the finite set of sampling points.
Such a discrete approximation to the expansion coefficients $a_{lm}$
is in fact  exact for all functions $f$ that are exactly equal to a
finite series of spherical harmonics up to order $l_{\mathrm{max}}$
if: i) the angles $(\theta_{k},\phi_{k})$ are chosen as the roots of
the basis functions of order $l = l_{\mathrm{max}} + 1 $, and ii)
the $w_{k}$ are appropriate weights. In the present case
$l_{\mathrm{max}}=2$. Thus the special points are the 18 roots of
the spherical harmonics $Y_{l,m} (\theta,\phi)$ of order $l=3$.
Furthermore, one may appreciate that the spherical harmonic
transform is a simple Fourier transform over $\phi$, and a Legendre
transform over $\cos \theta$. The weights $w_k$ are hence the
weights appropriate for Gauss-Legendre quadratures of order 3.
Explicitly, the 18 special points occur at azimuthal angles $\phi =
m\pi/3$ ($m=0,1,\ldots 5$) and at polar angles $ \theta
=\arccos(\sqrt{3/5}), \pi/2, \arccos(-\sqrt{3/5})$. The weights $w$
only depend on $\theta$, and are $5/9$ for $\theta=\arccos(\pm
\sqrt{3/5})$ and $8/9$ for $\theta=\pi/2$. Since one half of the 18
points  (see figure~\ref{specialpoints}) is antipodal to the other
half, inversion symmetry of the emission rate means that the rate
need only be evaluated for 9 dipole orientations in order to find
the full spherical harmonic expansion.

\section{Results for several nanophotonic examples}

\begin{figure*}
\includegraphics[width=\textwidth]{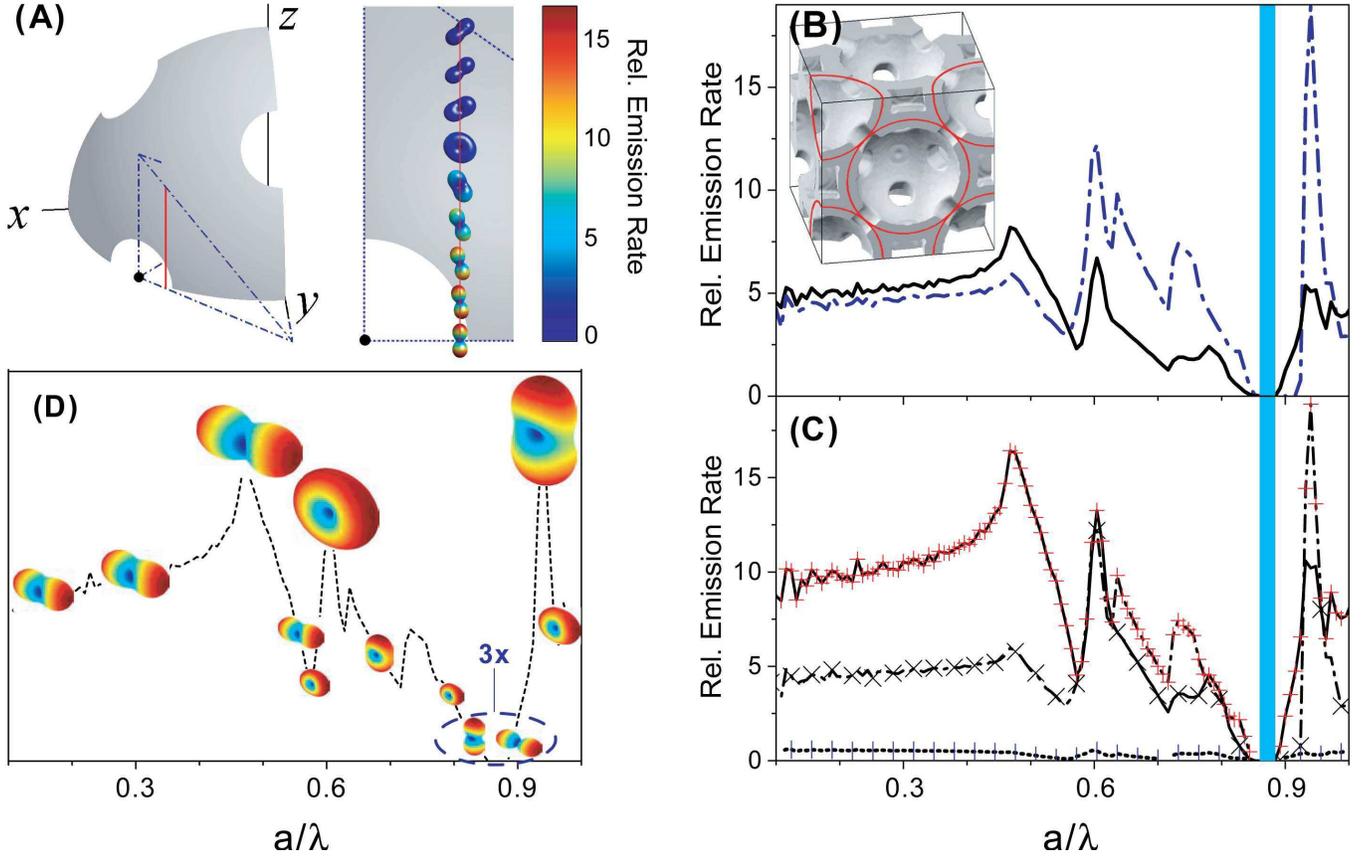}
\caption{\label{GammaFreqSiWindow} (color) Emission rate for a
quantum emitter in a photonic bandgap crystal. \textbf{(A)} Left:
$1/8$th of a cubic unit cell, blue dashed lines delimit the
primitive cell. Right: emission-rate surfaces on $x=0.2$, $y=0.3$,
variable $z$ (red line in left panel) at reduced frequency
$a/\lambda= 0.94$ ($a$ is lattice parameter). Surfaces are colored
by relative rate (scalebar on right), and have constant size.
\textbf{(B)} Emission rate for dipole centred in a window of a Si
inverse opal [$\mathbf{r}=1/4(1,1,0)$, black dot in (A)] with
orientations $\mathbf{e}_d = (1,0,0)$, $(0,1,0)$ (black curve), and
$(0,0,1)$ (blue dashed-dotted curve) versus $a / \lambda$. The rate
is normalized to the one in vacuum. The blue vertical bar indicates
the photonic bandgap. Inset: cubic unit cell. \textbf{(C)} Maximum,
medium, and minimum emission rates $\Gamma_{max}$ ($+$),
$\Gamma_{med}$ ($\times$), $\Gamma_{min}$ ($|$) compared to rates
for orientations $\mathbf{e}_d = (-1,1,0)/\sqrt{2}$ (full curve),
$(0,0,1)$ (dashed-dotted), $(1,1,0)/\sqrt{2}$ (short dots).
\textbf{(D)} Emission-rate surfaces at select frequencies show
strong changes in shape. The size of the surfaces is in proportion
to the absolute emission rates, and colorscales range from
$\Gamma_\mathrm{min}$ to $\Gamma_\mathrm{max}$. Dashed curve:
$\Gamma_\mathrm{max}$.}
\end{figure*}

To illustrate our analysis, we discuss the emission dynamics of a quantum emitter inside a photonic crystal, illustrated in
Fig.~\ref{GammaFreqSiWindow}(A,B). These complex systems have
extreme variations of the emission rate versus frequency on account
of a bandgap where emission is completely inhibited~\cite{Yab87}. To
obtain the rate for an emitter of arbitrary orientation in a Si
inverse opal, we have calculated the LDOS for the 9 special
orientations by summing over all Bloch eigenmodes~\cite{noteBloch}.
The crystal has a first order `pseudogap' at reduced frequency 0.55,
and a photonic bandgap from 0.852 to 0.891.
Figure~\ref{GammaFreqSiWindow}(B) shows the emission rate for a
salient position in the unit cell (\emph{cf.}
Fig.~\ref{GammaFreqSiWindow}(A)): the rate is anisotropic for
frequencies near the pseudogap, since it differs for dipoles
pointing in either $x,y$ or the $z$ direction, which are the cubic
symmetry axes of the crystal. One might be tempted to perceive the
behavior to be as simple as a mirror, since it is the same for both
$x$ and $y$. However, a plot of the maximum, medium, and minimum
emission rates (Fig.~\ref{GammaFreqSiWindow}(C)) shows that this
perception is completely wrong: Already at low frequency up to the
pseudogap, the emission rate is strongly anisotropic. While
anisotropic behavior in the long-wavelength limit may seem
surprising, its origin in electrostatic depolarization effects has
been discussed before~\cite{Miyazaki98,Rogobete03}. The maximum rate
occurs for dipole orientation $\mathbf{e}_d = (-1,1,0)/\sqrt{2}$,
and is much larger than the rate for any of the $x,y,z$
orientations, whereas the minimum rate for $\mathbf{e}_d =
(1,1,0)/\sqrt{2}$ is much smaller. At high frequency ($a/\lambda >
0.6$) up to the bandgap, the orientation of maximum rate changes to
$\mathbf{e}_d = (0,0,1)$. While it is clear from
Fig.~\ref{GammaFreqSiWindow}(C) that the orientation dependent
emission rate is much more complex than expected from (B),
Fig.~\ref{GammaFreqSiWindow}(C) hardly gives an intuitive picture of
the orientation-dependent behavior.

Therefore, we plot in Figure~\ref{GammaFreqSiWindow}(D) emission-rate surfaces versus frequency. At frequencies below the
pseudogap, the emission-rate surface is peanut-like, revealing that
the emission rate is high for a "horizontal" dipole orientation, and
inhibited for the 2 perpendicular orientations. At the pseudogap,
the emission-rate surface suddenly changes to donut-like, since the
rate is high for two orientations and low for a third orientation.
At even higher frequencies, the emission-rate surface becomes again
peanut-like - with donut-like behavior near 0.8 - but with a
different orientation than below the pseudogap. The maximum emission
rate is up to 20-fold enhanced, and the anisotropy
($\Gamma_{\mathrm{max}}/\Gamma_{\mathrm{min}}$) is strong with peaks
up to 340. In this particular example, the high symmetry at this
spatial position fixes all principal axes. To demonstrate the
applicability of our method to general, nonsymmetric, cases we have
also studied low-symmetry positions at constant frequency, see
Fig.~\ref{GammaFreqSiWindow}(A). Again strong anisotropies occur,
with the maximum-emission axis (or inhibition-axis) continuously
changing direction as a function of source position. We conclude
that emission-rate surfaces provide a compact representation of the
rich behavior of the dependence of the emission rates on dipole
orientation.

\begin{figure*}
\includegraphics[width=\textwidth]{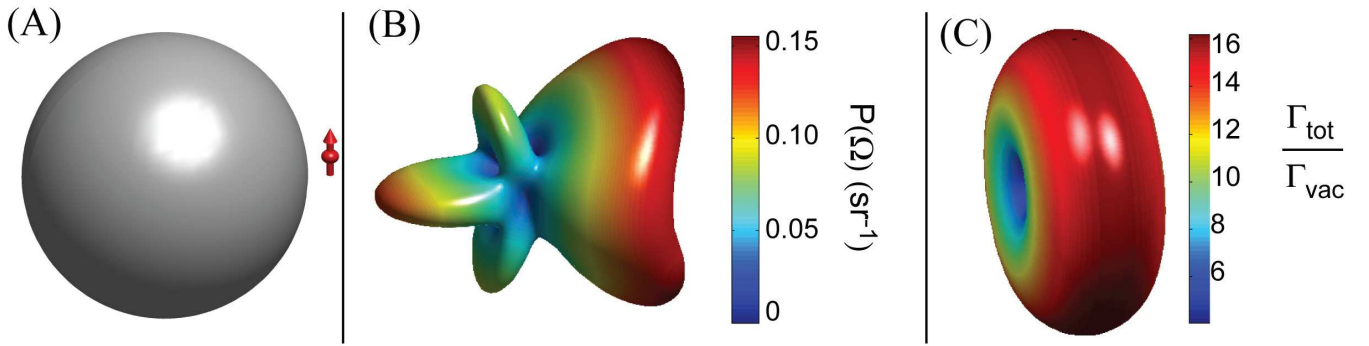}
\caption{\label{GammaPlasmonSphere} (color) \textbf{(A)} Drawing of
a two-level quantum emitter at 20 nm distance from a plasmonic Ag
sphere~\cite{palik} with radius $R=80$ nm. \textbf{(B)} Angular
distribution of the radiated power versus solid angle $\Omega$ for a
single dipole orientation parallel to the surface; the pattern has a
complex five-lobed structure. \textbf{(C)} Emission-rate surface
showing the emission rate versus dipole orientation. The pattern has
a donut-like shape.}
\end{figure*}

We emphasize that our classification of emission dynamics by means of emission-rate surfaces is by no means restricted to dielectric systems and can also be applied to dissipative nanophotonics systems that are of modern interest, such as plasmonic and metamaterial structures. Our analysis rests purely on the symmetry of the Green
dyadic in Eq.~(\ref{LDOS2Green}), which in the presence of optical
absorption describes the total decay rate (radiative rate plus
induced nonradiative rate) of a quantum emitter. As an example, we
discuss the textbook case of an emitter near a plasmonic
sphere~\cite{Anger06,Kuhn06}, using the known Green's function~\cite{Taibook} (\emph{cf.}
Fig.~\ref{GammaPlasmonSphere}(A)).
Figure~\ref{GammaPlasmonSphere}(C) shows that the emission-rate surface for the total decay rate has a donut-like shape
($\Gamma_{\mathrm{min}} \leq \Gamma_{\mathrm{med}} =
\Gamma_{\mathrm{max}}$) with 16-fold enhanced rates for a dipole
parallel to the surface, and 5-fold enhanced for a perpendicular
dipole. For a fixed dipole orientation~\cite{Novotny06,Kuhn2008},
the angular distribution of the radiated power reveals a well-known
five-lobed structure (B). A comparison of (B) and (C) illustrates
the main differences between radiation patterns and emission-rate
surfaces: radiation patterns are relevant to a \textit{single}
dipole orientation and do not necessarily have any symmetry, or are
free to follow any symmetry inherent in the environment.
Emission-rate surfaces on the other hand are relevant to all
orientations and have a symmetry limited by the quadratic form.

\section{Discussion}
Since the analysis in this paper is based on $\mbox{Im}(\mathbb{G}(\mathbf{r},\mathbf{r}))$ it is strictly valid for the \emph{total} decay rate modification induced by the nanophotonic environment. Explicitly, in the case of losses our proof only holds for the sum of the radiative rate and the non-radiative rate ($\Gamma_\mathrm{rad}+\Gamma_\mathrm{nonrad}$), and
not for the radiative rate $\Gamma_\mathrm{rad}$ separately. To analyze the radiative emission-rate surfaces one would need to analyze the far-field integral of the radiated power (quantity in Fig.~\ref{GammaPlasmonSphere}(B)) as a function of the source
orientation. A priori it is not at all clear that such radiative rate surfaces need have a quadratic form. Indeed, we have not succeeded in proving the quadratic form for the radiative rate in the lossless case by analysis of far-field integrals, \emph{i.e.}, \emph{without\/} identifying $\Gamma_{\mathrm{rad}}=\Gamma_{\mathrm{tot}}$ and subsequently analyzing $\mbox{Im}(\mathbb{G}(\mathbf{r},\mathbf{r}))$. We have numerically calculated radiative emission rate surfaces for many low-symmetry dissipative plasmon sphere clusters, and have not found any example in which the radiative emission rate surface was not quadratic. Although a rigorous proof is beyond the scope of this paper, we therefore anticipate that the quadratic form not only holds for total decay rates, but also for radiative decay rates.

A class of quantum emitters with a single transition dipole moment
are fluorescent molecules, such as laser dyes~\cite{Novotny06}. For
such emitters, emission-rate surfaces can be observed if their
orientation is controlled, \textit{e.g.}, by attaching them to
liquid crystal molecules that are oriented in external fields
\cite{Gottardo06}. If one can tune the orientation of an emitter,
this opens a novel opportunity to "switch" spontaneous emission from
inhibited to enhanced and \textit{vice versa}. The emission-rate
surfaces reveal that optimal switching always requires a dipole
rotation by $90^\circ$, since minimal and maximal emission rates
always occur along the mutually perpendicular main axes.
Alternatively, one could tune semiconductor nanowires with oriented dipole moments. For self-assembled and colloidal quantum dots with dipoles in a $x',y'$ plane, we expect to probe the $x',y'$ cross-sectional average of the emission-rate surface of the relevant nanophotonic system.

Since arbitrary orientations do not usually coincide with principal
dipole orientations, most prior work on specific systems has been
incomplete, since no principal rates has been reported. While such
incompleteness does not affect the orientation averaged rate (see
Appendix~\ref{Averages}), it does affect the understanding of dynamics of orientational dipole
ensembles~\cite{Gerard98,Lodahl04,Muskens07}. Such a decay is a sum
of single exponentials with a rate distribution given by the emission-rate surface. Any observable derived from time-resolved
decay beyond the orientation-averaged rate ($\mathrm{Tr}(\mathrm{Im}(\mathbb{G}))$) requires knowledge of the
principal rates, which is thus relevant to many physical situations in nanophotonics.

In classical optics, the imaginary part of the Green dyadic is not
only relevant for radiating dipoles. Indeed, the imaginary part of
the Green dyadic has also been connected to the so-called coherency
matrix (or the electric cross-spectral density
tensor)~\cite{MandelWolf95} for black body radiation. In general,
the $3\times 3$ coherency matrix describes second-order spatial
correlations of the electric field, and can be understood as a
generalization of Stokes parameters to quantify the polarization of
near fields locally. Within this framework, a description of local
polarization by polarization ellipsoids directly points at a
quadratic form of the coherency matrix, since ellipsoids are
\emph{level sets} (rather than polar plots) of an equation of the
form in Eq.~(\ref{decayratebasis}). It should be noted that the
coherency matrix depends on the incident source that generates the
local electric field. In the particular case that the field is due
to black body radiation the coherency matrix reduces to the
imaginary part of the Green dyadic
$\mbox{Im}(\mathbb{G}(\mathbf{r},\mathbf{r}))$, as derived by
Set\"{a}l\"{a} \emph{et al.}~\cite{Setala03}. However, it is
important to realize that for this identification of
$\mbox{Im}(\mathbb{G}(\mathbf{r},\mathbf{r}))$ with the coherency
matrix to hold, the source is required to be a statistically
homogeneous and isotropic distribution of radiating currents, and
the medium is supposed to be non-dissipative~\cite{Setala03}. This
is diametrically opposite to the analysis of spontaneous emission
sources presented here, which concerns localized and oriented
sources and is valid without limitation on material dissipation. It
is exciting that our method to find principal rates and orientations
can be directly adapted to calculate the local polarization
properties of black body radiation.

\section{Summary}
We have theoretically studied the spontaneous emission rate of a
two-level quantum emitter in any nanophotonic system. We derive a
general representation of the dependence of emission rates on the
orientation of the transition dipole by only invoking symmetry of
the Green function. The rate depends quadratically on orientation
and is determined by rates along three principal axes. We show that
these principal rates and axes can be easily calculated without
evaluation of the full Green function. Furthermore we show that
visualization of emission-rate surfaces as determined from principal
rates provides great insight on how preferred orientations for
enhancement (or inhibition) depend on emission frequency and
location, and on strategies to actively switch emission rates by the
dipole orientation, as shown for a mirror, a plasmonic sphere, or a
photonic bandgap crystal.

\section{Acknowledgments}
We thank Allard Mosk, Ad Lagendijk, Peter Lodahl for useful discussions.
This work is part of the research program of the Stichting voor Fundamenteel Onderzoek der Materie (FOM) that is financially supported by the Nederlandse Organisatie voor Wetenschappelijk Onderzoek (NWO).
WLV also thanks NWO-Vici and STW/NanoNed.

\appendix
\section{Discussion of average emission rate} \label{Averages}
A remarkable fact is that the orientation-average rate $\langle
\Gamma \rangle$ can \emph{always} be calculated from the LDOS at
just three perpendicular orientations, which need not coincide
with the principal axes
$\{\mathbf{v}_\mathrm{min},\mathbf{v}_\mathrm{med},\mathbf{v}_\mathrm{max}\}$.
First, we calculate the orientation averaged rate by integration
over the full emission surface. Without loss of generality we
align $x,y,z$ with the principal axes, so that the
orientation-dependent rate is:
\begin{equation}\label{GammaPrincip}
\Gamma(\theta,\phi) = \Gamma_{\mathrm{min}} \cos^2 \phi \sin^2
\theta + \Gamma_{\mathrm{med}} \sin^2 \phi \sin^2 \theta +
\Gamma_{\mathrm{max}} \cos^2 \theta.
\end{equation}
By straightforward integration, the orientation-averaged rate
$\langle \Gamma \rangle$ is
\begin{equation}\label{GammaAverage}
\langle \Gamma \rangle= \frac{1}{4\pi} \int_0^{2\pi} d\phi
\int_0^\pi \Gamma(\theta,\phi) \sin\theta d\theta =
\frac{1}{3}(\Gamma_{\mathrm{min}}+\Gamma_{\mathrm{med}}+\Gamma_{\mathrm{max}}).
\end{equation}
Integration over the full emission surface clearly shows that the
orientation-averaged emission rate is equal to the mean of the three
principal rates, and hence $\langle \Gamma \rangle= (\pi
d^2\omega/\hbar\epsilon_0)(2\omega/\pi c^2\cdot
\mathrm{Tr}(\mathrm{Im}(\mathbb{G}(\mathbf{r},\mathbf{r},\omega)))$.
The invariance of the trace of any matrix under arbitrary basis
rotation implies that the average rate in Eq.~(\ref{GammaAverage})
can be calculated from the rates at any randomly chosen but mutually
orthogonal directions $x, y, z$ as
\begin{equation}\label{GammaAverageXYZ}
\langle \Gamma \rangle =
\frac{1}{3}(\Gamma_{x}+\Gamma_{y}+\Gamma_{z}).
\end{equation}

\printfigures

\end{document}